# HOW UNUSUAL IS THE LOCALLY QUIET HUBBLE FLOW?

David Schlegel, Marc Davis and Frank Summers
Astronomy Department, University of California, Berkeley, CA 94720

AND

Jon A. Holtzman
Lowell Observatory, Mars Hill Road, 1400 West, Flagstaff, AZ 86001
*Written 1993 October 21*

## ABSTRACT

The local galaxy distribution offers an interesting constraint to cosmological models of structure formation. The galaxies are distributed in a long, filamentary structure, presumably the result of large amplitude gravitational instability, yet the local velocity field is cold. In particular, there are no blueshifted galaxies within $5h^{-1}$ Mpc except those within the Local Group radius of $1h^{-1}$ Mpc. Using numerical simulations we demonstrate that such a situation is extremely rare for an observer in $\Omega = 1$ CDM models, but is not uncommon in mixed dark matter (MDM) models that include massive neutrinos.

*Subject headings:* cosmology: theory – dark matter – galaxies: clustering – large-scale structure of the universe

## 1. THE CHALLENGE OF THE LOCAL GROUP

Our local neighborhood ($r \lesssim 5h^{-1}$ Mpc) is the best locale in the Universe for gauging the "temperature" of the clustered galaxy distribution. The local velocity field is observed to be cold, yet the clustering is well developed. This is indicative of a low mass/light ratio if the distribution is in virial equilibrium. The problem is well known (e.g. de Vaucouleurs 1975, Tully and Fisher 1987) and has recently been emphasized by Peebles (1992) as a "grand challenge" for the proponents of high density, hierarchical models of structure formation.

This paper is a direct response to Peebles' challenge. We attempt to demonstrate that the standard CDM model has severe difficulties creating the observed, cold velocity field. On the other hand, we demonstrate via numerical simulation that a model with both hot and cold dark matter (Mixed Dark Matter, or MDM), normalized to match the COBE fluctuations on large scales, has little difficulty in producing filamentary chains that match both the morphology and cold velocity field of our local environment. The present study is meant to be an existence proof: that our quiet local environment does not preclude the possibility that we live in a high density universe.

## 2. OBSERVATIONAL DATA AND CONSTRAINTS

The local galaxy distribution is plotted in redshift space in Figure 1. Galaxies within a box with sides equal to 800 km/s, oriented with respect to supergalactic coordinates, have been selected from Huchra's ZCAT (1992). (We shall quote all distances in this paper in units of km/sec to avoid confusion with the Hubble constant.) We have excluded only those galaxies that are obviously members of Virgo, whose high peculiar velocities would have scattered them into picture. All 36 galaxies with a projected distance of less than three degrees from RA=$12^h$28.3, DEC=+12°40 have been excluded. The existence of the supergalactic plane is striking in the SGY-SGZ plot. If the galaxies trace the mass distribution without too much bias, our neighborhood is collapsed on scales of $8h^{-1}$ Mpc. The Milky Way resides in a pancake or filament structure that extends in the direction of Virgo, but also extends on the opposite side of the sky. This collective assembly is labeled the Coma-Sculptor cloud (e.g., Fisher and Tully 1987).

Deviations from Hubble flow, or peculiar velocities, can be derived only when distances are known independently, allowing one to subtract the Hubble expansion of each galaxy. Accurate distances are available from infrared Tully-Fisher measurements for 270 galaxies and groups of galaxies within $cz = 4200$ km/s (Aaronson et al. 1982; Pierce & Tully 1992; as presented in Shaya et al. 1992 and Tully et al. 1992). Figure 2 plots the most local of these distances versus radial velocities taken from ZCAT, corrected to the Local Group frame using the best solution from Yahil et al. (1977): $v_\odot = 308$ km/s toward $l = 105°$, $b = -7°$. Tully et al.'s distances have been converted to km/s using $H_0 = 88$ km/s/Mpc. The straight line represents the expected modulus for pure Hubble flow. The 15 plotted points represent 31 individual galaxies, where some have been collapsed into small groups by Tully et al. Individual distances were not available, although none have negative velocities (blueshifts).

Observational errors, estimated to be 16% in distance, have contributed to the scatter of the points in the horizontal direction; the redshift errors are negligible. The remaining deviations represent radial peculiar velocities. Even assuming the distance measurements are perfect, the scatter in this plot is surprisingly small. The dispersion in radial velocity of all 270 points is 429 km/s from the best-fit straight line. The dispersion of the 67 points within 1000 km/s is 236 km/s, while the dispersion of the 15 points within 500 km/s is a mere 84 km/s. Correcting for distance errors via quadrature subtraction, the instrinsic rms peculiar velocity in this very local sample of 15 points is only 60 km/s.

A fundamental constraint can be made without resorting to any distance measures. In our selected volume of space there are no blueshifted galaxies outside of the Local Group, a fact that has been argued to constrain the local velocity dispersion to be as low as 50 km/s (Sandage & Tammann 1975, and Fisher & Tully 1975). Only in the core of the Virgo cluster are there galaxies with sufficient peculiar motion to appear blueshifted.

One final observational constraint is the local galaxy or mass overdensity relative to the mean Universe. The IRAS 1.2 Jy survey (Fisher et al. 1992) is perhaps the best catalog to compute this overdensity. It samples the full-sky Universe out to distances of at least 10000 km/s with well-defined selection criteria. We have computed the number density of IRAS galaxies that would be brighter than 1.2 Jy at a distance of 500 km/s, using the procedures described in Yahil et al. 1991. We find the mean density is 0.060 galaxies per $(h^{-1}$ Mpc$)^3$ for any windows larger than 6000 km/s. The local density within top-hat windows of radius 500-1000 km/s is never below 0.075 per $(h^{-1}$ Mpc$)^3$, corresponding to a local overdensity of 0.25. Strauss et al. (1992) argued that the IRAS galaxies faithfully trace the optical galaxy distribution everywhere except in the cores of rich clusters. The error in the IRAS density estimate is dominated by the uncertainties in the slope of the faint end of the luminosity distribution function, and is slightly sensitive to corrections from redshift space to real space. Our estimate is derived in redshift space, except we have collapsed clusters to prevent contamination from the likes of Virgo. Any attempt to select regions from simulations that are statistically similar to the Local Group region should match this local overdensity. It is difficult to quote an error to this overdensity estimate, but at the very least we are in an overdense region ($\delta > 0$). The observational constraints can be summarized as follows:
(1) The local galaxy distribution is filamentary, and the local overdensity, averaged within a sphere of radius 500 km/s, is approximately 0.25.
(2) No galaxies outside the Local Group (100 km/s) and within 500 km/s are blueshifted.
(3) The local velocity field is "cold".

### 3. SIMULATIONS

We have run several high-resolution N-body simulations with the intent of identifying observers that satisfy these observations in a gross sense. We have chosen to study the standard cold dark matter (CDM) model and the mixed (hot plus cold) dark matter model (MDM), both with $H_0 = 50$ km/s/Mpc and $\Omega = 1$. Previous tests of the MDM model using simulation techniques have been discussed by Davis, Summers & Schlegel (1992) and Klypin et al. (1993). Only dark matter particles were used in the simulations, with cold particles accounting for 70% and hot particles 30% of the mass in the mixed model. The CDM transfer function was from Bardeen et al. (1986), and the MDM transfer functions were from Holtzman (1992). Both models use the same random numbers when creating the initial conditions via the Zel'dovich approximation, so the same regions of space can be compared. The mass normalizations are chosen to match an rms signal of $13.6\mu$K on $10°$ scales, consistent with the COBE measurement (Smoot et al. 1992). This corresponds to $\sigma_8 = 1$ for the CDM model, and $\sigma_8 = 2/3$ for the MDM model. Earlier outputs can be studied from the CDM simulation which correspond to lower values of $\sigma_8$. We shall focus on the $\sigma_8 = 2/3$ output of the CDM simulation in order to provide a direct and more favorable comparison to the MDM model.

In the MDM simulation, we include the physics of both dark matter components. Different transfer functions were necessary for each component since the neutrinos erase their initial fluctuations on small scales due to free-streaming. Particle positions were perturbed from a grid using the standard Zel'dovich approximation. The bulk velocities of both the cold particles and neutrinos were generated from the time derivatives of the transfer functions, which takes account of the suppressed growth rate at high wavenumbers. In addition, the neutrinos are each assigned a random, primordial velocity drawn from a Fermi-Dirac distribution that has adiabatically decayed to

$$f(v) = \frac{1}{1 + e^{(v/v_0)}}$$

with a characteristic velocity $v_0 = 7.2(1+z)$ km/s for neutrinos with a mass of 7.0 eV. One neutrino was used for each cold particle, with a mass ratio of 3 : 7. We have found this sampling of neutrino phase space to be adequate. Denser sampling is a waste of particles if the initial amplitudes are not set too low (see Klypin et al. 1993 for a contrary view). We find that changes in the random numbers for the primordial neutrino velocities only slightly perturb the later structure formation.

The simulations were evolved using a P$^3$M N-body code. We used $64^3$ particles (per dark matter species) in a periodic box with a comoving size of 2500 km/s, evolved from the epoch $(1+z) = 11$. For these parameters, the particle mass is $3.3 \times 10^{10} M_\odot$ in the CDM simulation. The cold particles and neutrinos are $2.3 \times 10^{10} M_\odot$ and $1.0 \times 10^{10} M_\odot$, respectively, in the MDM simulation. The force softening parameter was set to $\epsilon = 0.001$ of the box size, which is sufficient to resolve galaxies and galaxy groups. "Galaxies" were identified using a friends-of-friends algorithm with the linking parameter $b = 0.15$, corresponding to overdensities of $\delta \sim 1/b^3 \sim 300$. We chose to study only those galaxies with a minimum mass of 5 particles, corresponding to $1.7 \times 10^{11} M_\odot$. The results are insensitive to this number, but this choice gives a number density that approximates the observed number density of galaxies. There are approximately 3000 galaxies identified in each simulation, for a number density of 0.2 per $(h^{-1}$ Mpc$)^3$. This is a 'galaxy' density a factor of three smaller than shown in Figure 1, which includes many faint, dwarf galaxies, but is sufficient for our purpose.

Figure 3 (a,b) is a plot of a slice 800 km/s thick of all particles in the simulation boxes for the CDM and MDM models at $\sigma_8 = 2/3$, while figure 3 (c,d) is a plot of the identified galaxies in these slices. Some of the galaxies are actually groups of galaxies artificially merged by our identification procedure, but this is equivalent to the merging that has occurred for the real data in Figure 2. The periodic box has been recentered on filament B (see below). Note that the "natural bias" mechanism (White et al. 1987) has, as expected, depopulated the low density regions rather effectively. The bias is small: on scales where $\xi = 1$, the galaxies are biased relative to the cold matter by a factor $b \sim 0.9$ in the CDM model and $b \sim 1.2$ in the MDM model, although this measure is sensitive to the small scale merging. Note also that the MDM model has remarkably straight filaments with relatively little substructure, while the filaments in the CDM model are largely broken into dense knots.

## 4. ANALYSIS

We ask the question of how successful each of the models is in providing Local Group candidate positions that satisfy the observational constraints discussed in section 2. In the simulations, it is straightforward to identify which galaxies have no infalling neighbor galaxies in the range of 100 to 500 km/s. A large fraction of the galaxies satisfy this single criterion in all models: $\sim 15\%$ in CDM at any $\sigma_8 \leq 1$ and $\sim 25\%$ in MDM at $\sigma_8 = 2/3$.

As expected, these "no infall" galaxies are selected preferentially in low-density regions, mainly because the low density regions are not dynamically relaxed. Densities have been computed from the galaxy counts in top-hat spheres of radius 500 km/s. These numbers can be directly compared to the local density measured by the IRAS survey. Restricting oneself to higher density regions, one is left with far fewer candidate observers. In fact, the CDM simulation has only one or two galaxies, of approximately 3000, at any redshift that satisfy both the "no infall" criterion and $0.25 < \delta < 0.5$. The MDM model, with both suppressed growth and suppressed velocities on small scales ($r <$ few Mpc), fares much better; our MDM simulation has 55 galaxies that satisfy both criteria. In figure 3 (c,d), the galaxies satisfying each criterion are identified. Figure 4 shows the distribution of galaxies with no blueshifted neighbors as a function of galaxy density. The sparseness of Local Group candidates does not change as the CDM model is evolved from low amplitude to high amplitude. In the MDM model, most galaxies would have neighbors with blueshifts, but nearly half of the galaxies in regions where $\delta = 0.25$ have no infalling neighbors. Thus constraint number 2 is not statistically improbable for galaxies that adequately meet the first, overdensity constraint. These fractions are listed in Table 1.

One expects that the MDM models would create quieter velocities for a given initial fluctuation amplitude, or $\sigma_8$, than CDM or even tilted CDM models. The peculiar velocities of the cold component are somewhat suppressed by the free-streaming of the hot component, as has been discussed by Davis *et al.* (1992), an effect not present in the other models.

To address constraint number 3, we have selected several candidate positions for more detailed analysis of the velocity field. We have chosen regions by eye that seem to generally satisfy the morphology criterion that our Local Group is in the midst of a filamentary or pancake-like structure. Location A is in an obvious pancake structure with a large mass overdensity, $\delta = 2.0$ within 500 km/s. Location B is in a filamentary structure satisfying our criteria in the MDM simulation ($\delta = 0.3$). Location C is centered on the only galaxy in the CDM simulation at $\sigma_8 = 2/3$ that satisfies criteria 1 and 2 ($\delta = 0.3$).

High density filaments such as location A are unlikely Local Group candidates. The "no infall" criterion is not satisfied for any galaxies in the region. The velocity dispersions are much too high for any of the models, as evidenced in Figures 6. The lower-density regions, such as B and C (see Figures 5 and 6), are much more successful at matching the observed velocity field. In such regions, the MDM model successfully produces a cold velocity field. The rare Local Group candidates in the CDM simulation still have very high velocity dispersions.

Of the 55 MDM candidates satisfying the "no infall" criterion and with $0.25 < \delta < 0.5$, the average, radial velocity dispersion is 138 km/s for all galaxies within 500 km/s. For the only candidate in the CDM model at $\sigma_8 = 2/3$, the dispersion is 185 km/s. For the only candidate in the CDM model at $\sigma_8 = 1$, the dispersion is 258 km/s. One could relax the density constraint somewhat and insist upon $0 < \delta < 0.5$: the CDM model is still anemic in Local Group candidates that match our constraints at any $\sigma_8$ normalization. These results are summarized in Table 1. The 1D velocity dispersions about the Local Group candidates should be compared directly to the observed value of 60 km/s, as discussed in section 2.

The morphology of the regions in the MDM models populated with Local Group candidates are usually filamentary. These are collapsed structures that are often only slightly overdense in a spherical average compared to the mean. Consequently it is no surprise that they are not collapsing along the remaining primary axis, which would generate a number of blueshifted neighbors. In fact, these filamentary structures seem to be draining to other more overdense regions. This is supported by an apparent, local Hubble flow that is slightly larger than the mean Hubble expansion (see Figure 6). The more overdense structures continue to collapse upon themselves. The pancake structure at Location A is such a case, where the velocity field is hot and local flow is retarded.

## 5. CONCLUSIONS

The fact that we see no blueshifts outside of the Local Group and the Virgo cluster argues strongly against the possibility that we live in anything more than a slightly overdense region of the Universe, at least in highly evolved CDM and MDM cosmologies. This is consistent with the results from the IRAS survey, $\delta \sim 0.25$ in a sphere of radius 500 km/s around us, and would be only slightly lower if galaxies are a biased tracer of mass.

Filament B in the MDM simulation has a "galaxy" distribution that is remarkably similar to the local environment seen in Figure 1. Many of these regions have no blueshifted neighbors and yet have local densities $\delta > 0$. The neighboring voids are more depleted in "galaxies" than mass, but in fact the bulk of the mass has migrated to the higher density zones. No model of random phase structure formation will fully deplete the low density regions for any value of $\Omega$.

In the CDM model, it is rare to find Local Group candidate regions that simultaneously satisfy the first two of our three observational constraints. It is even rarer to locate CDM candidates simulataneously satisfying all three of our constraints. The MDM models are much more successful at creating Local Group candidates that agree with the observed local overdensity and cold velocity field. Nearly half of the Local Group candidate galaxies in our MDM model with $\delta = 0.25$ have no blueshifted neighbors, so the phenomenon is not rare in this model. Thus the MDM model remains a very promising theory of structure formation.

Unfortunately, the local group observations are a statistic of one event; it will be some time before peculiar velocities are sufficiently accurate to allow us to play this game in other locales.

REFERENCES

Aaronson, M. et al. 1982, ApJS, **50**, 241
Bardeen, Bond, Kaiser, & Szalay 1986, ApJ, **304**, 15
de Vaucouleurs, G. 1975, Galaxies & the Universe, Stars and Stellar Systems, Vol 9, eds. A. Sandage, M. Sandage & J. Kristian, (Chicago: Univ. of Chicago Press), 557.
Davis, M., Summers, F.J., & Schlegel, D. 1992, Nature, **359**, 393
Fisher, J.R. & Tully, R.B. 1975, A&A, **44**, 151
Fisher, K.B. 1992, Ph.D. thesis (Univ. of California at Berkeley)
Holtzman, J.A. 1992, private communication
Huchra, J.P. 1992, private communication
Klypin, A., Holtzman, J., Primack, J., & Regös, E., 1993, ApJ, in press
Peebles, P.J.E. 1992, Relativistic Astrophysics and Particle Cosmology, Texas/PASCOS 92 Symposium, eds. C.W. Akerlof & M.A. Srednicki, (New York: New York Acad. Sci.), Vol 688, 84
Pierce, M.J. & Tully, R.B. 1992, ApJ, **387**, 47
Sandage, A. & Tammann, G.A. 1975, ApJ, **196**, 313
Schlegel, D. et al., in preparation
Shaya, E.J., Tully, R.B. & Pierce, M.J. 1992, ApJ, **391**, 16
Smoot, G.F. et al. 1992, ApJ, **396**, L1
Strauss, M.A., Davis, M., Yahil, A. & Huchra, J.P. 1992, ApJ, **385**, 421
Tully, R.B., & Fisher, J.R., Nearby Galaxies Atlas, Cambridge, 1987
Tully, R.B., Shaya, E.J. & Pierce, M.J. 1992, ApJS, **80**, 479
Yahil, A., Strauss, M.A., Davis, M., & Huchra, J.P. 1991, ApJ, **372**, 380
Yahil, A., Tamman, G.A., & Sandage, A. 1977, ApJ, **217**, 903
White, S.D., Davis, M., Efstathiou, G. & Frenk, C.S. 1987, Nature, **330**, 451

**Table 1.** Comparison of the observations with the models.

| | Observations | MDM $\sigma_8=2/3$ | CDM $\sigma_8=0.45$ | CDM $\sigma_8=2/3$ | CDM $\sigma_8=1.0$ |
|---|---|---|---|---|---|
| No. of galaxies in simulation box † | | 2657 | 3320 | 3206 | 2879 |
| $f_{NB}(-1<\delta<\infty)$* | | 26% | 14% | 16% | 16% |
| $f_{NB}(0.25<\delta<0.5)$ | | 27% | 0.4% | 0.3% | 0.4% |
| $f_{NB}(0<\delta<0.5)$ | | 30% | 5% | 3% | 1% |
| Average 1D velocity dispersion ‡ | 60 km/s | 138 km/s | 138 km/s | 185 km/s | 258 km/s |

† $H_0 L=2500$ km/s

* $f_{NB}(\delta)$ is the fraction of galaxies at density $\delta$ having no blueshifts

‡ within 500 km/s of the galaxies having no blueshifted neighbors and $0.25<\delta<0.5$

FIGURE CAPTIONS

Fig 1. Galaxies in a cube with sides 800 km/s in redshift space, oriented with respect to supergalactic coordinates. The 36 galaxies within three degrees of Virgo have been excluded, leaving 290 galaxies in this plot.

Fig 2. Recession velocities versus derived Tully-Fisher distances for galaxies and small groups of galaxies with distances less than 500 km/s. Velocities have been converted to the Local Group frame.

Fig 3. Slices of the cold dark matter distribution in the simulations that are 2500 km/s wide and 800 km/s thick. The different panels are for (a) particles in the CDM simulation at $\sigma_8 = 0.68$, (b) particles in the MDM simulation at $\sigma_8 = 0.68$, (c) galaxies from the CDM model, and (d) galaxies from the MDM model. Galaxies with no blueshifted neighbors in the range 100 km/s to 500 km/s are circled. Of these galaxies, those that reside in overdense regions ($\delta > 0$ within $R = 500$ km/s) are filled circles. The inner square identifies location B, shown in redshift space in Figure 5.

Fig 4. Distribution of galaxies with no blueshifted neighbors as a function of galaxy density. The dotted line is the density distribution of all galaxies. The different panels are for (a) CDM at $\sigma_8 = 0.68$ and (b) MDM at $\sigma_8 = 0.68$.

Fig 5. Three selected regions in the simulations in redshift space. The different panels are for (a) CDM at $\sigma_8 = 0.68$ and (b) MDM at $\sigma_8 = 0.68$.

Fig 6. Hubble plots for the selected observer positions in Figure 5 using all galaxies within $5h^{-1}$ Mpc of those observers. The dispersion in these plots should be compared to the observational data presented in Figure 2.

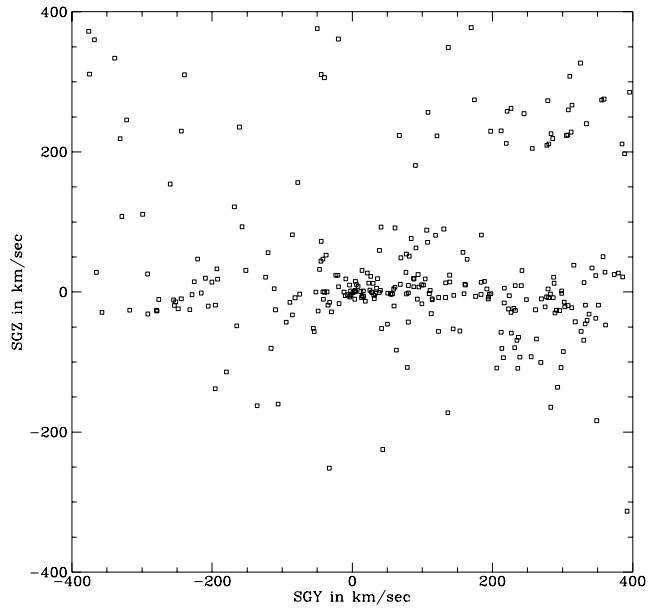

Fig 1. Galaxies in a cube with sides 800 km/s in redshift space, oriented with respect to supergalactic coordinates. The 36 galaxies within three degrees of Virgo have been excluded, leaving 290 galaxies in this plot.

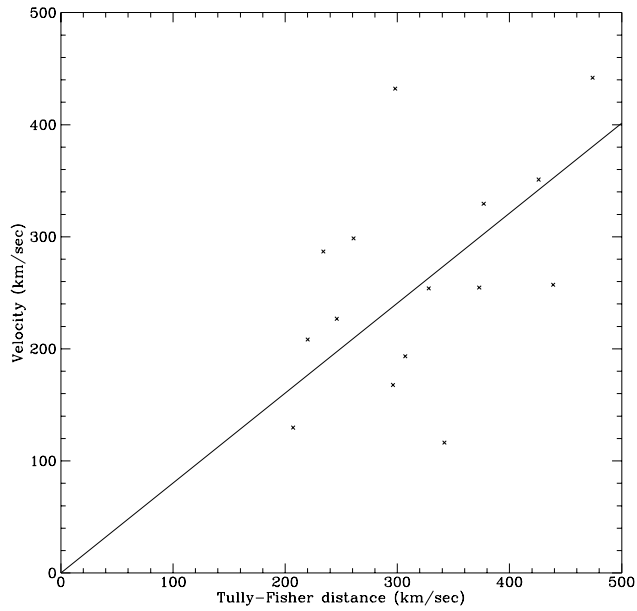

Fig 2. Recession velocities versus derived Tully-Fisher distances for galaxies and small groups of galaxies with distances less than 500 km/s. Velocities have been converted to the Local Group frame.

FIGURES 3A AND 3B AVAILABLE UPON REQUEST
VIA ANONYMOUS FTP TO COMA.BERKELEY.EDU

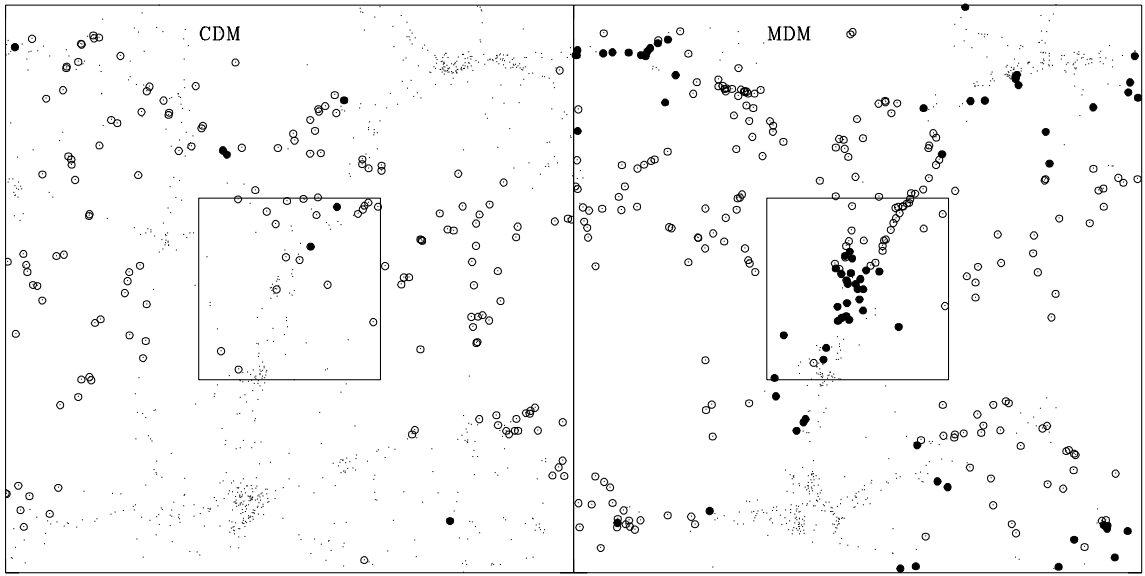

Fig 3. Slices of the cold dark matter distribution in the simulations that are 2500 km/s wide and 800 km/s thick. The different panels are for (a) particles in the CDM simulation at $\sigma_8 = 0.68$, (b) particles in the MDM simulation at $\sigma_8 = 0.68$, (c) galaxies from the CDM model, and (d) galaxies from the MDM model. Galaxies with no blueshifted neighbors in the range 100 km/s to 500 km/s are circled. Of these galaxies, those that reside in overdense regions ($\delta > 0$ within $R = 500$ km/s) are filled circles. The inner square identifies location B, shown in redshift space in Figure 5.

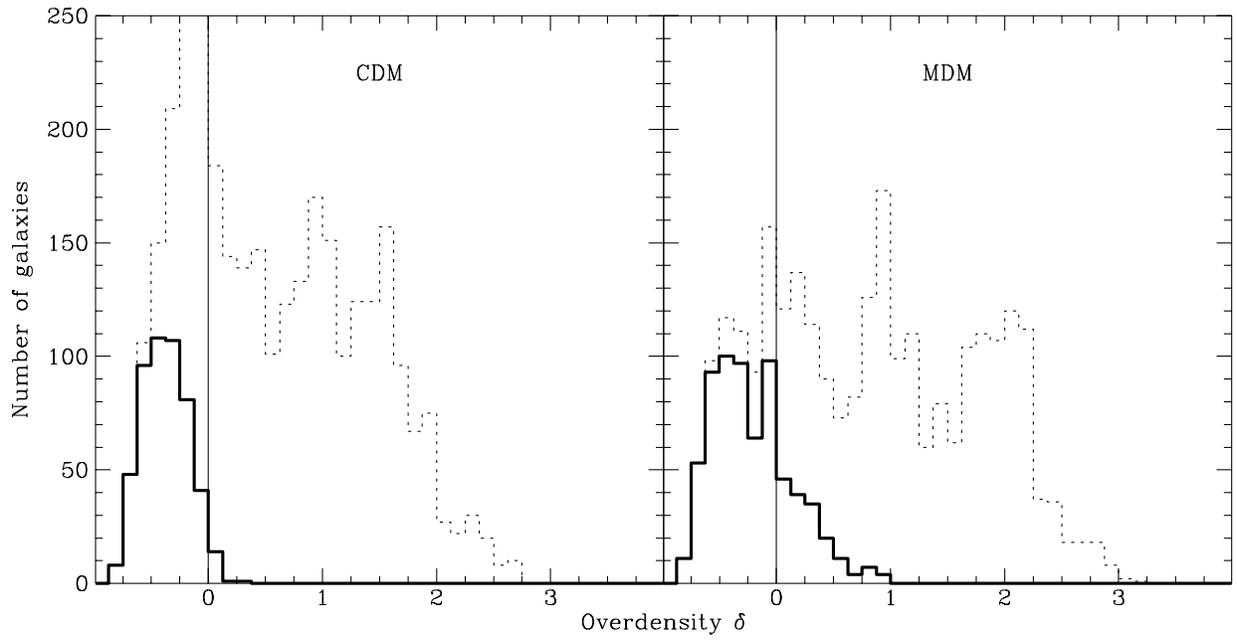

Fig 4. Distribution of galaxies with no blueshifted neighbors as a function of galaxy density. The dotted line is the density distribution of all galaxies. The different panels are for (a) CDM at $\sigma_8 = 0.68$ and (b) MDM at $\sigma_8 = 0.68$.

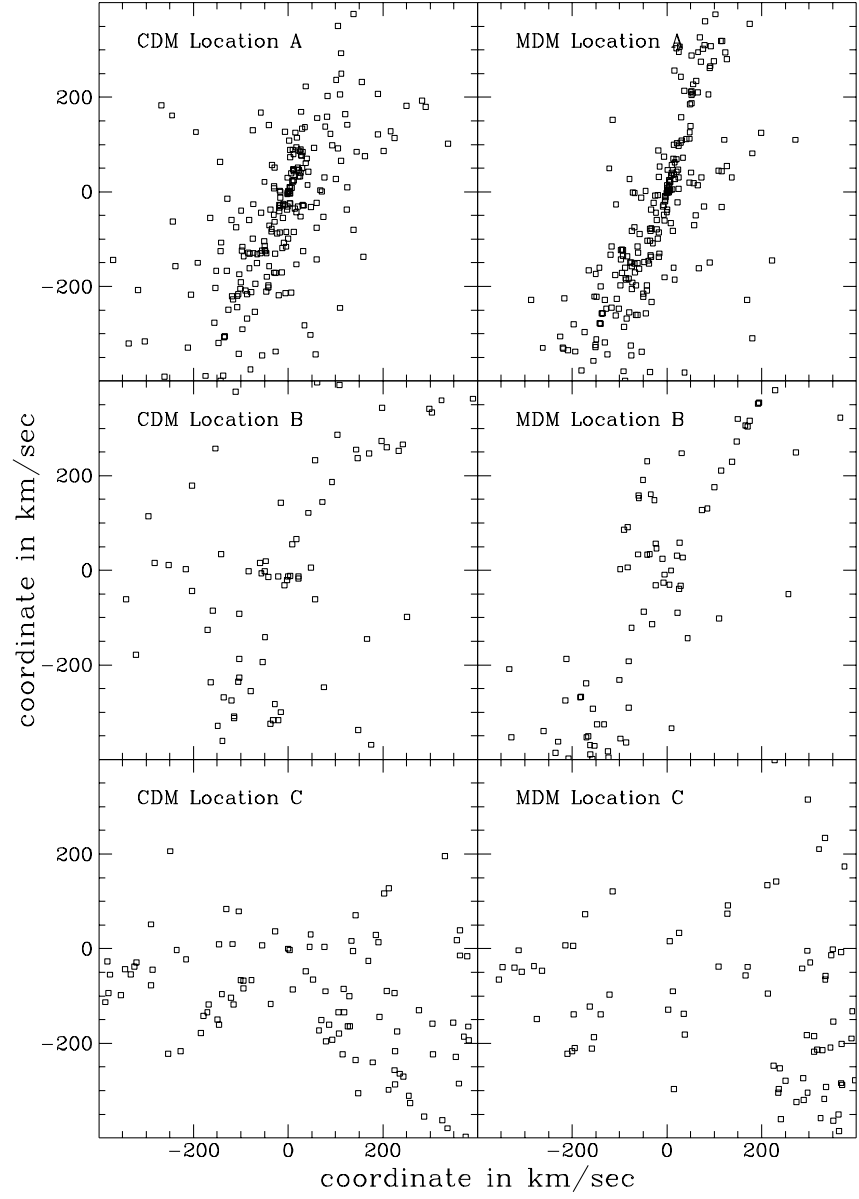

Fig 5. Three selected regions in the simulations in redshift space. The different panels are for (a) CDM at $\sigma_8 = 0.68$ and (b) MDM at $\sigma_8 = 0.68$.

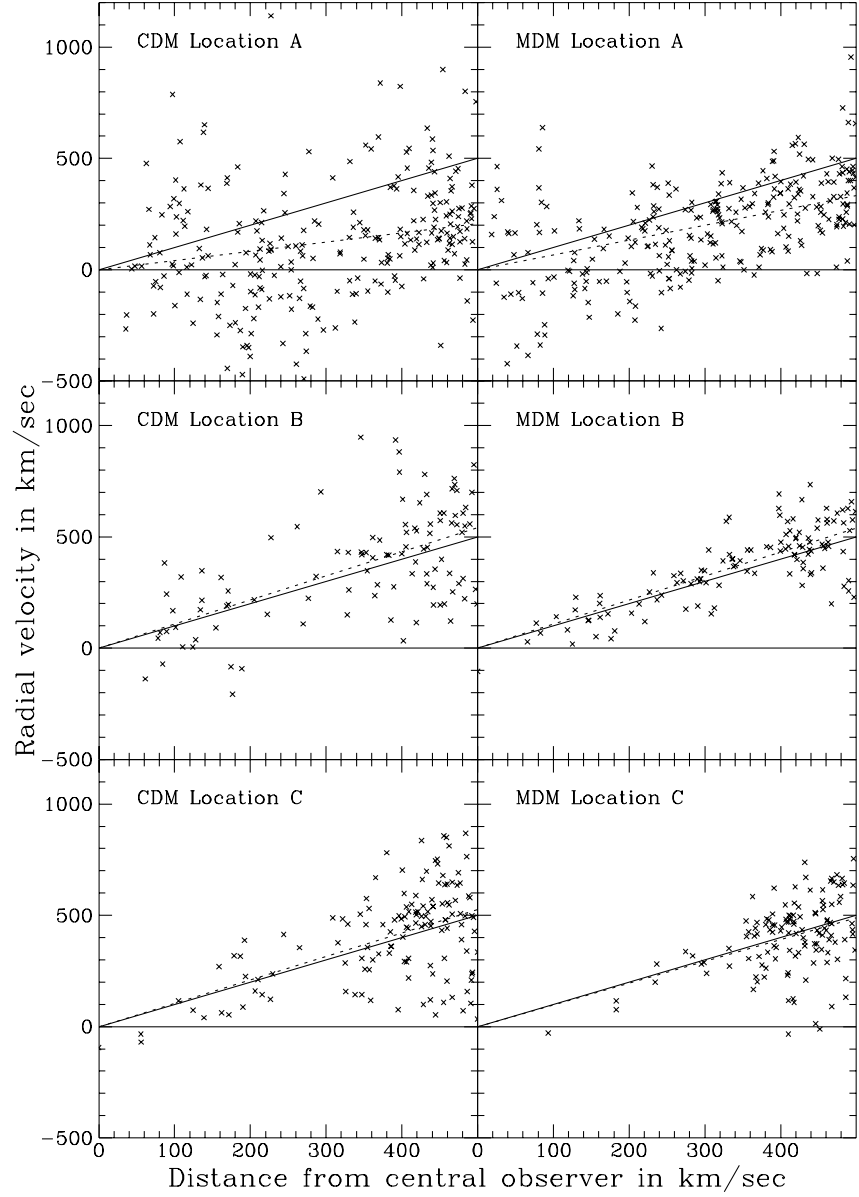

Fig 6. Hubble plots for the selected observer positions in Figure 5 using all galaxies within $5h^{-1}$ Mpc of those observers. The dispersion in these plots should be compared to the observational data presented in Figure 2.